\documentstyle[epsfig,aps,twocolumn,prl]{revtex}
\begin{document}

\newcommand{\ket}[1]{| #1 \rangle}
\newcommand{\bra}[1]{\langle #1 |}
\newcommand{\braket}[2]{\langle #1 | #2 \rangle}
\newcommand{\proj}[1]{| #1\rangle\!\langle #1 |}

\draft
\title {Evolutionary quantum game}
\author{Roland Kay, Neil F. Johnson and Simon C. Benjamin}
\address {Physics Department and Center for Quantum Computation,
\\ Clarendon Laboratory, Oxford University, Parks Road, Oxford, OX1 3PU,
U.K.}
%
\date{\today}
\maketitle


\begin{abstract}  We present the first study of a dynamical quantum game.
Each agent has a  `memory'
of her performance over the previous $m$ timesteps, and her strategy can
evolve in time.
The game exhibits
distinct regimes of optimality. For small $m$ the classical game
performs better, while for intermediate $m$ the relative performance depends
on
whether the source of
qubits is `corrupt'. For large
$m$, the quantum players dramatically outperform the classical players by
`freezing'
the game into high-performing attractors in which evolution ceases.

\end{abstract}
\vskip\baselineskip

\vskip\baselineskip



The new field of quantum games is attracting significant
interest\cite{eisert,meyer,simon1,xu}.  We recently conjectured
\cite{private} that novel features should arise for quantum games with
$N\geq 3$ players. Benjamin and Hayden \cite{simon2}
subsequently created a Prisoner's Dilemma-like game for $N=3$ with a
high-payoff
`coherent quantum equilibrium' (CQE). Johnson\cite{neil} showed
that this quantum advantage can become a disadvantage when
the game's external qubit source is corrupted by a noisy `demon'. So far,
all
studies have been restricted to static games.

Here we introduce an iterated version of the game, where players (agents)
may
modify their strategies based on information from the past -- i.e. they
`learn' from
their mistakes
\cite{usprl}.
This evolutionary game
produces highly non-trivial dynamics in both quantum and classical regimes,
and represents
the first step toward understanding iterated games employing temporal
quantum
coherence. Agents are provided  with the minimum resources necessary for
adaptability -
specifically, each agent posesses a measure of her past success through a
parameter
$\$_m$ whose value reflects the payouts from recent rounds of the
game. The fixed `memory' parameter $m$ effectively governs the number of
rounds upon which $\$$
depends (see later Eq. (1)), and turns out to be fundamental for deciding
the
relative superiority of the quantum and classical games.  For small memory
$m$, the
classical game is superior (in the sense that the average payout per player
is higher). For
intermediate
$m$,  relative superiority is determined by the reliability of the external
qubit
source, while for large
$m$ the quantum game is radically superior due to evolutionary `freezing'
into a high-paying
attractor.

\begin{figure}
\centerline{\epsfig{file=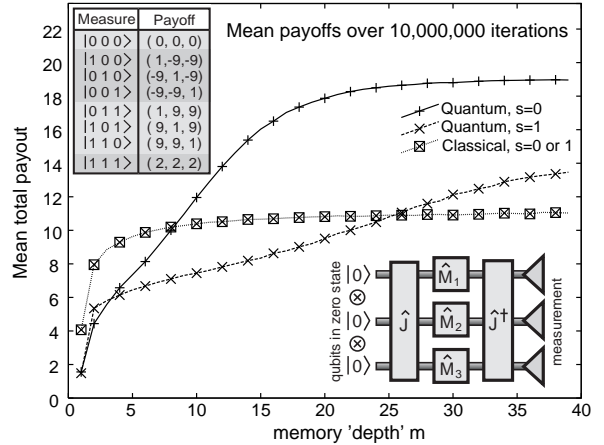,width=7.8cm}}
\vspace{0.2cm}
\caption{Mean total payout per turn as a function of the memory size $m$.
Curve with crosses in
boxes: classical game with input qubit source $0$ or $1$. Curves with
crosses: quantum games with
input qubit source $|0\rangle$ and $|1\rangle$ respectively. Upper inset:
Payoff table for our
three-player game. Lower inset: circuit diagram showing flow of information
in the
general quantum game. When the
$J$ gates are removed, the game becomes classical.}
\end{figure}

We briefly review the static version of the game \cite{simon2,neil}.
The basic idea is to establish the flow of information in a game, and then
investigate
the consequences of quantizing this information.
In the classical static  game, each of the $N=3$ players receives a bit
in the zero state, and chooses either to `flip', or `not flip' this bit
prior to returning it.
The three bits are then measured, and the payoff to the players is
determined by the table
shown in Fig. 1. This payoff table is such that
each player's best choice of strategy is simply to choose to `flip' {\em
regardless} of what any
other player may be doing. Given that all players realize this, the
inevitable outcome is
that all players choose `flip' - this is the game's `dominant strategy
equilibrium' (DSE). The
final measured state is therefore
$111$, and the corresponding payout to each player is
just
$2$ points. This result arises despite the existence of cooperative
strategies, such as
each player choosing `flip' with probability 80\%, which would yield a
higher (expected)
payoff to all players: the problem is that any given player is better off
switching strategy to
`definitely flip' hence this cooperative strategy profile does not occur.
This is
exactly the same defection problem seen in the famous game Prisoner's
Dilemma.
The quantization process \cite{simon2} involves: (a) generalizing
each bit to a qubit $\alpha\ket{0}+\beta\ket{1}$, (b) generalizing the moves
available
to the players from `flip'/`don't flip' to a set of quantum operations, and
(c) introducing `gates' to entangle the qubits - specifically the entangling
gate is ${\hat
J}=\frac{1}{\sqrt 2}({\hat I}^{\otimes 3} + i {\hat F}^{\otimes 3})$ where
${\hat F}={\hat
\sigma}_x$. Without entanglement, the quantum game is trivially equivalent
to the classical
variant \cite{simon2} - therefore we can make a correspondence between the
quantum and classical
games simply by removing the $J$ gates. The lower inset in Fig.1 shows
the flow of information in the quantum game. In general the actions of the
players
in the quantum game will result in a final state which is a superposition.
Measurement then
collapses this state to one of the classical outcomes, and the payoff is
then be determined
from the table (see Fig. 1) as in the classical game. Among the conclusions
of Ref. \cite{simon2}
was the discovery that the quantum players could escape the DSE that traps
the classical
players, and hence outperform the classical players by a considerable
factor. In Ref. \cite{neil}
it was shown that this quantum advantage depends on the reliability of the
source of qubits: if
the source is believed to be generating qubits in the $\ket{0}$ state, but
is in fact generating
qubits in state $\ket{1}$, then the classical game players will out-perform
the quantum ones.

Following the previous literature on iterated classical multi-player games
\cite{usprl}, we limit the classical moves to three options:
`definitely flip', `definitely do not flip' and `flip with probability 1/2'.
These options are
denoted by setting $p$, the probability of leaving the input qubit
unflipped, equal
to $0$,$1$ or $1/2$ respectively. This simplification from the full
(continuous) range of
physically possible $p$ values clarifies the analysis while retaining the
basic structure of the
game. There are now
$3^3=27$ possible profiles or `configurations' $(p_1,p_2,p_3)$. These yield
ten
`classes' each containing $C\geq 1$ configurations which are equivalent
under
interchange of player label\cite{usprl}. The upper table in Fig. 2 shows the average
payoffs for
each configuration class for the classical game.  Given that the input is 0,
the dominant
strategy equilibrium corresponds to all players choosing
$p=0$, i.e. class (iv) in the table. Hence although the continuous-parameter
$p$-space has been discretized to only three values, this description
includes
the fundamental dominant strategy equilibrium.

To maintain a correspondence between the quantum and classical games, we
also permit our
quantum players just three different moves: $\hat I$,
$\hat\sigma_x$, and
$\frac{1}{\sqrt 2}(\hat\sigma_x + \hat\sigma_z)$. It was earlier shown that
these moves can give
rise to superior quantum performance for the static quantum game with
reliable
qubit source \cite{simon2}. Moreover when the entangling
$J$ gates are removed, making the game classical, these three moves
correctly correspond to our
allowed classical moves
$p=1,0,1/2$ (respectively). Therefore we label the three quantum moves with
the same
$p$ notation. A fully comprehensive study of the quantum game would also
examine
the case where the players are permitted a larger set of moves, e.g. one
that is closed under
composition. However the aim here is to make a first direct move-for-move
comparison between the
iterated quantum and classical games, and this necessitates restricting the
larger set of quantum
options.  The resulting table (see Fig. 2) provides a simple quantum analog of the
classical case.

\begin{figure}
\centerline{\epsfig{file=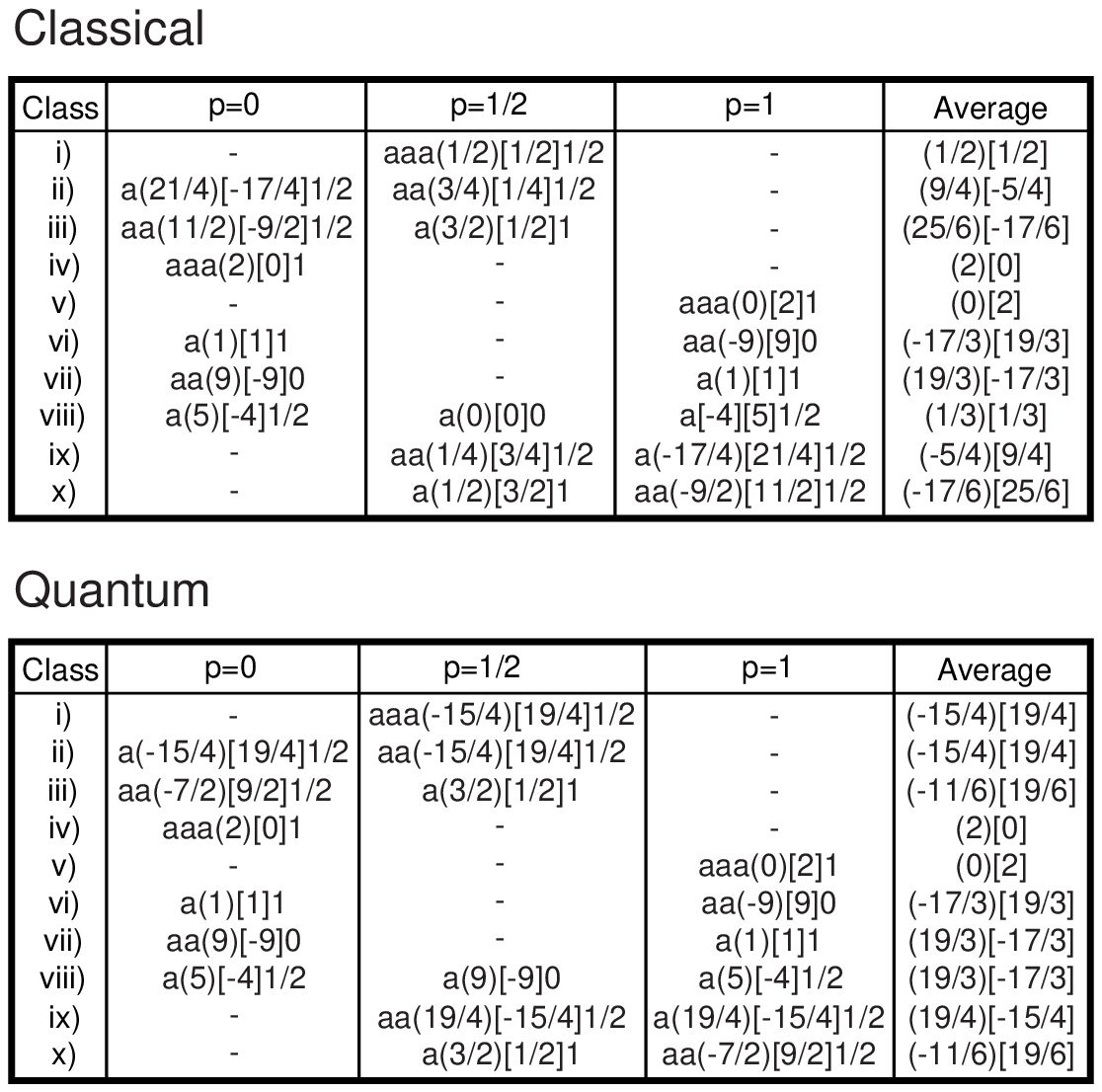,width=7.8cm}}
\vspace{0.2cm}
\caption{Average payoffs for classical game (upper)  and quantum game
(lower).
Agents are denoted by `a'.
$p$ corresponds to the probability of not flipping the input qubit - for the
quantum case, $p$
is an operator (see text).  Average payoffs for input qubit 0 are shown as
$(\dots)$; those for input 1 are shown as $[\dots]$;
those for
$50:50$ mixture of input qubits are shown without parentheses.
Final column gives the average payoff per player for a given input qubit.}
\end{figure}

We assume that in both quantum and classical games,
players are unable to communicate between themselves and hence cannot
coordinate which player picks which strategy. In the quantum game, this is
more
critical since the CQE (i.e. the Nash equilibrium given by class (viii) in
the lower table of Fig. 2) involves players using different
$p$'s. We will find that the evolutionary aspect of our iterated game allows
the quantum players to find cooperative solutions even in the absence of
communication, provided
that the memory $m$ is not too small.

The agents are allowed to evolve (`mutate') their
strategies based on past success, according to the following prescription.
After timestep $t-1$, each player holds a
`moving average' $\$_{m,t-1}$ of her payoffs which is then updated at
timestep $t$:
\begin{equation}
\$_{m,t} = \frac{m-1}{m}\$_{m,t-1} + \frac{1}{m}\Delta_t
\end{equation}
where $\$_{m,t}$ is the updated moving average and
$\Delta_t$ is the payoff to the player at turn $t$. The information about
previous outcomes
therefore has a `half life' since the contribution of a given round's payoff
falls exponentially with successive rounds. When the player's moving average
falls below a certain threshold $d$, she mutates - she chooses one of the
other two
allowed strategies with a $50:50$ chance. After mutation, the moving average
$\$_m$ continues to
be updated, but the player enters a `trial period' of $Y$ turns before
considering
mutating again. For simplicity, we set $Y=m$ for the remainder of this
paper. We also consider
fixed external qubit source orientations, although each of these
restrictions can be relaxed. We
choose a mutation threshold
$d=3.0$ in order to be above the $2$ point maximum possible payoff per
player in the classical
static game (see payoff table in Fig. 1). We performed runs with $10^7$
timesteps - this is
sufficient for the resulting statistics to be stable to within a few
percent.

Figure 1 shows the results for the average payoff per
turn for the three agents combined. The classical dominant strategy
equilibrium
$(2,2,2)$ corresponds to a value 6, while the CQE $(5,9,5)$ and its
permutations
correspond to a value 19. Figure 3 shows the
corresponding average lifetime of the agents before mutation.
Figure 1 shows that for memory $m<8$, the quantum game players are {\em
worse} off than
the classical game players regardless of the input qubit source orientation
(0 or
1). (The particular orientation of the fixed source doesn't affect the
results for the classical game.) The classical game players are also less
prone
to mutation, as seen from Figure 3. For $8<m<25$, the quantum game players
do
better than the classical players if the source is 0 (as they believe it is)
but do worse if the
source is 1 (i.e. the source is `corrupt' \cite{neil}). The same is
qualitatively true for the
lifetimes (Fig. 3) but for a smaller range
$8<m<19$. For $m>25$, the quantum game players achieve better average
payoffs
than the classical game players irrespective of the source orientation, but
those with the reliable (i.e. non-corrupt) source do significantly better.
They eventually
`freeze' into a cooperative state, as explained later.

\begin{figure}
\centerline{\epsfig{file=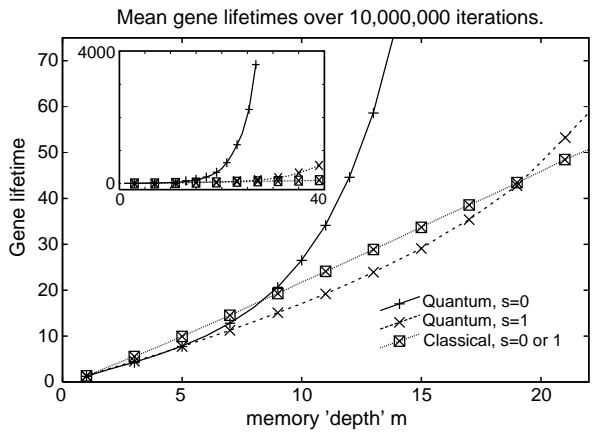,width=7.8cm}}
\vspace{0.2cm}

\caption{Average time agents maintain a given strategy before mutating, as a
function of the
memory $m$. Inset: Results for a larger y-scale, illustrating the `freezing'
of the
dynamics in the quantum game with a reliable qubit source ($s=0$).}
\label{figure1}
\end{figure}

Figure 1 reveals an interesting effect within the classical game: for $m>1$
the
classical game players do better than the classical dominant strategy
equilibrium value of
$2+2+2=6$. If we had set the threshold $d\leq 2$, the players would quickly
`freeze' into the configuration $p=0,0,0$, which guarantees 2
points per player per turn. By adopting a higher satisfaction limit, i.e. by
being
more `greedy', the classical players introduce instability into the game --
although this
implies that in any given round at least one player is typically worse off,
in the long term the
instability actually benefits all the players.

The explanation for the interesting large $m$ behavior lies in the cycles
through which the game
moves on short and long timescales. In general, the behavior of both
classical and quantum
systems for larger
$m$ is dominated by a small number of possible attractors which trap the
system for a certain period, before it jumps into a brief period of
turbulence and then ends up in another attractor.

\begin{figure}
\centerline{\epsfig{file=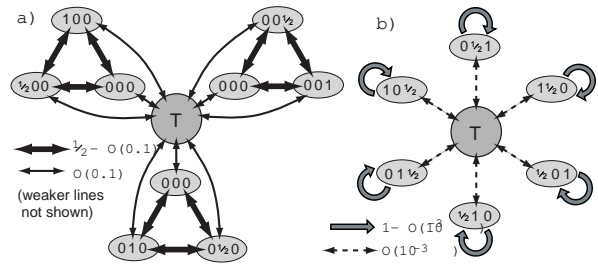,width=7.8cm}}
\vspace{0.2cm}
\caption{Schematic game dynamics showing the attractors which dominate both
the classical and
quantum games in the large
$m$ limit. (a) The classical game, and (b) the quantum game with a reliable
source ($s=0$). Ellipses represent specific strategy profiles and are
labeled by the
$p$ values adopted by the three players. The circle labeled $T$ represents
the set of all other
profiles, which are visited only during brief turbulent periods in the
game's dynamics. Arrows
indicate the probability that one profile follows another over a period of
$m$ rounds. Data are
collected over
$10^7$ rounds with $m=30$. In the classical game the profile
$(p_1=0,p_2=0,p_3=0)$, corresponding to the dominant strategy equilibrium,
occurs in each of the
three attractor cycles.}
\end{figure}

Figure 4 shows the structure of the attractors in the classical and quantum
cases. In the
classical game, the attractors correspond to two players fixing their
strategy as `definitely
flip' ($p=0$). While they maintain these choices, the third player can never
achieve satisfactory
payouts, hence the short timescale dynamics correspond to this third player
jumping
around the three possible
$p$ values, constantly mutating. Eventually one of
the other two players will encounter a run of losses and mutate her
strategy. The game
then enters a (poorly performing) turbulent period before settling back into
either the
previous attractor or one of the 2 permutation-equivalent cycles. In the
quantum case, however,
these three short cycles are replaced by six attractors, each a single
configuration such as
$p=0,1,1/2$. Because all such attractors yield satisfactory expected payoff
to all players, the
system will remain in an attractor until one player has an exceptionally
long run of losses.
Figure 3 indicates that the probability of this occurence falls
exponentially with $m$ - i.e. the
system `freezes' into one configuration profile.

In summary, we have presented the first results for an evolutionary quantum
game.
The dynamics are non-trivial, even in the present case where the game's
history is assumed
to be classical in that a measurement is taken at each timestep. The next
logical step is to
to allow for a superposition (or `multiverse') of histories over $R>1$
turns. Because
the coherence in the present game need only be maintained for $R=1$ turns,
the present
game could be implemented as an algorthim in an elementary quantum computer
containing
$3$ or more qubits - such an experiment could be performed with existing NMR
technologies.  More generally, it is not inconceivable that such `games' are
already being played
at some microscopic level in the physical world  - indeed, it has recently
been
shown that nature plays classical dominant-strategy games
using clones of a virus that infects bacteria\cite{turner}.


\end{document}